\documentclass[aps,pra,floatfix,amsmath,preprint,amsshowpacs,11pt]{revtex4}
\usepackage{graphicx}
\begin{document}

\title{Shuffling cards; factoring numbers; and the quantum baker's map}
\author{Arul Lakshminarayan}
\affiliation{Department of Physics\\ Indian Institute of Technology Madras\\
Chennai, 600036, India.}

\preprint{IITM/PH/TH/2005/3}

\begin{abstract}
It is pointed out that an exactly solvable permutation operator, viewed as
the quantization of cyclic shifts, is useful in constructing a basis
in which to study the quantum baker's map, a paradigm system of quantum
chaos. In the basis of this operator the eigenfunctions of the quantum
 baker's map are compressed by factors of around five or more.
We show explicitly its connection to an operator that is closely
 related to the usual quantum baker's map. 
This permutation operator has interesting connections to the art of shuffling cards
as well as to the quantum factoring algorithm of Shor via the quantum order finding
one. Hence we point out that this well-known quantum algorithm makes
crucial use of a quantum chaotic operator, or at least one that is
close to the quantization of the left-shift, a closeness that we also explore
quantitatively. 
\end{abstract}
\pacs{05.45.Mt,03.67.Lx}
\maketitle

\newpage

\newcommand{\newc}{\newcommand}
\newc{\beq}{\begin{equation}}
\newc{\eeq}{\end{equation}}
\newc{\kt}{\rangle}
\newc{\br}{\langle}
\newc{\beqa}{\begin{eqnarray}}
\newc{\eeqa}{\end{eqnarray}}
\newc{\pr}{\prime}
\newc{\longra}{\longrightarrow}
\newc{\ot}{\otimes}
\newc{\rarrow}{\rightarrow}
\newc{\h}{\hat}
\newc{\bom}{\boldmath}
\newc{\btd}{\bigtriangledown}
\newc{\al}{\alpha}
\newc{\be}{\beta}
\newc{\ld}{\lambda}
\newc{\sg}{\sigma}
\newc{\p}{\psi}
\newc{\eps}{\epsilon}
\newc{\om}{\omega}
\newc{\mb}{\mbox}
\newc{\tm}{\times}
\newc{\hu}{\hat{u}}
\newc{\hv}{\hat{v}}
\newc{\hk}{\hat{K}}
\newc{\ra}{\rightarrow}
\newc{\non}{\nonumber}
\newc{\ul}{\underline}
\newc{\hs}{\hspace}
\newc{\longla}{\longleftarrow}
\newc{\ts}{\textstyle}
\newc{\f}{\frac}
\newc{\df}{\dfrac}
\newc{\ovl}{\overline}
\newc{\bc}{\begin{center}}
\newc{\ec}{\end{center}}
\newc{\dg}{\dagger}
\newc{\prh}{\mbox{PR}_H}
\newc{\prq}{\mbox{PR}_q}

A textbook example of a simple fully chaotic system is provided by the model of 
a baker mixing dough, the baker's map.
 The classical baker's map \cite{LL}, $T$, is the area preserving
 transformation
of the unit square $[0,1)\times [0,1)$ onto itself, which takes a
phase space point $(q,p)$ to $(q',p')$ where $(q'=2q,\, p'=p/2)$ if
$0\le q<1/2$ and $(q'=2q-1,\,p'=(p+1)/2)$ if $1/2\le q<1$.  The
stretching along the horizontal $q$ direction by a factor of two is
compensated exactly by a compression in the vertical $p$
direction. The repeated action of $T$ on the unit square leaves the phase
space mixed, this is well known to be a fully chaotic system that in a
mathematically precise sense is as random as a coin toss \cite{Orn}. The
area-preserving property makes this map a model of chaotic two-degree
of freedom Hamiltonian systems, and the Lyapunov exponent is
$\log(2)$ per iteration.  

As the classical baker's map is exactly solvable in many ways, including an 
explicit prescription for finding periodic orbits of any period, its 
quantization was sought as a simple model of quantum chaos. The baker's map as
 quantized by Balazs and Voros \cite{BalVor} has many nice features, including 
simplicity,  that make it ideal for this purpose and has been used extensively 
in studies of quantum chaos and semiclassical methods. It has also been 
experimentally implemented recently using NMR \cite{NMR}. The quantum baker's 
map, in the position representation, that we use here is:%
\beq \label{bvsbak} B=G_N^{(\f{1}{2},\f{1}{2})\, \dagger}\left( 
\begin{array}{cc}G_{N/2}^{(\f{1}{2},\f{1}{2})} &0\\0 & 
G_{N/2}^{(\f{1}{2},\f{1}{2})} \end{array}\right), \eeq %
where 
\beq
(G_{N})_{mn}^{(\alpha,\beta)}= \df{1}{\sqrt{N}}\,\exp[-2 \pi i 
(m+\alpha)(n+\beta)/N].\eeq
We require that
 $N$ be an even integer; Saraceno \cite{Saraceno} imposed anti-periodic boundary
  conditions ($\alpha=\beta=1/2$) that we use. In this case we drop the 
superscripts  indicating these phases. 
The Hilbert space is finite dimensional, the dimensionality $N$ being the 
scaled inverse Planck constant $(N=1/h)$, where we have used that the 
phase-space area is unity. The position and momentum states are denoted as 
$|q_n\kt$ and $|p_m\kt$,where $m,n=0,\cdots,N-1$ and the transformation function 
between these bases is the finite Fourier transform $G_N$ given above.
 
 The choice of anti-periodic boundary conditions fully preserves parity
  symmetry, here called $R$, which is such that $R|q_{n}\kt = |q_{N-n-1}\kt.$ 
  Time-reversal symmetry is also present and implies in the context of
  the quantum baker's map that an overall phase can be chosen such that the
momentum and position representations are complex conjugates: $G_N
\phi=\phi^{*}$, if $\phi$ is an eigenstate in the position basis. 
$B$ is an unitary matrix, whose repeated application is the quantum version
 of the full left-shift of classical chaos. There is a semiclassical trace 
formula, which, based on the unstable periodic orbits, approximates eigenvalues 
\cite{Almeida}.

   Despite the simplicity
 of the quantum baker's map, its solution in terms of exact spectra continues
 to be elusive. Recently we showed \cite{meen} that for $N$ that are powers of 
two,  it is possible to write approximate analytic formulae for certain class of 
states. In particular the Thue-Morse sequence ($\{1,-1,-1,1,-1,1,1,-1,\ldots 
\}$, the $n$-th term is the parity of $n$ when expressed in the binary, counting 
$n$ from zero) \cite{Allpaper} and its Fourier transform \cite{Luck} determined 
to a large extent  a class of states we called ``Thue-Morse states''. Similar 
expressions were also found for families of strongly scarred states. Despite 
having simple, if approximate, analytic formulae these states were found to be 
multifractals. Thus we went some way in solving a quantum chaotic system that is 
nearly generic. A crucial tool used was the Walsh-Hadamard transform \cite{Schroeder}. 
That is, if $\phi$ is an eigenstate we studied $H_K \phi$, where $H_K=\otimes^K 
H$,  a $K$-fold tensor product of the Hadamard matrix 
$H=((1,1),(1,-1))/\sqrt{2}$, where $2^K=N$.
 
We wish to now address the case of 
general $N$ and arrive at a counterpart of the Walsh-Hadamard transform that 
will simplify the states of the quantum baker's map.   We show that a simple 
operator, the shift operator, that is exactly solvable, acts as a good zeroth 
order operator for the quantum baker's map. Therefore its eigenstates form a 
basis in which the eigenstates of the quantum baker's map  appear simple. We 
study this operator's action in phase space, and show how to build  a quantum 
baker's map around this operator. This ``new'' quantum baker will then turn out 
to be very close to the ``usual'' quantum baker's map in Eq.~(\ref{bvsbak}).   

The shift operator $S$, by 
definition, acts on the position basis as $S|q_n\kt = |q_{2n}\kt$ or 
$|q_{2n-N+1}\kt$ depending on if $n< N/2$ or otherwise. We notice that $S$ is 
``almost"  $B$, only there is no momentum cut-off, as $\br p_m|B|q_n\kt = 
\sqrt{2}\br p_m|q_{2n}\kt$   for $n$ and $m$ both $\le N/2-1$. In fact $S$ is a 
generalization of what was proposed as the quantum baker's map by Penrose 
\cite{Penrose} for the case when $N=2^K$. In this case if the position state 
$|q_n\kt$ is denoted in terms of the binary expansion of 
$n=a_{K-1}a_{K-2}\cdots a_0$ then $S|a_{K-1}a_{K-2}\cdots 
a_0\kt=|a_{K-2}a_{K-3}\cdots a_0 a_{K-1}\kt$. It is easy to see that $S$ 
commutes with the parity operator $R$. However $S$ does not respect the usual 
time-reversal symmetry, relevant to the baker's map, namely  $G^{-1}_NS^*G_N \ne 
S^{-1}$. It does respect a "restricted" time-reversal symmetry in the   case 
when $N=2^K$, as $\hat{b}^{-1}\,S^{*}\, \hat{b} = S^{-1}$, where $\hat{b}$ is 
the bit reversal operator defined as $\hat{b} |a_{K-1}a_{K-2}\cdots 
a_0\kt=|a_{0}a_{1}\cdots a_{K-2} a_{K-1}\kt$. It is useful to rewrite the 
action of $S$ on the position basis (written simply as $|n\kt$) as
\beq
S|n\kt = 
|2n \, \mbox{mod}(N-1)\kt,
\eeq
with the caveat that $S|N-1\kt=|N-1\kt$, rather than $|0\kt$. This is not 
crucial as it affects only an one-dimensional invariant subspace.
 
 We point to two 
apparently unrelated contexts in which $S$ has already appeared. Firstly $S$ is 
closely related to the perfect ``riffle-shuffle" \cite{cards} used to randomize 
a deck of cards, to be more precise the "out-shuffle".  If for instance $N=8$ 
 cards were in a deck, it is split into two exact halves and the cards are then 
 interleaved. If the cards were numbered $0,1,2,3,4,5,6,7$, the out-shuffle 
brings it to $0,4,1,5,2,6,3,7$, which is easily verified to be the action 
$S^{-1}$. The {\it deterministic} chaos of this shuffling process forms the 
basis of certain card tricks. The perfect shuffle returns the deck to its 
original state after a few shuffles, we will see below that this is the 
``quantum period function'' relevant to $S$.   
 
 Secondly, a generalization of $S$, where the factor 2 is replaced by any 
integer (coprime to $N-1$)  is precisely the operator whose ``phase estimation'' 
leads to the solution of the   order-finding problem \cite{Shor}.
 The multiplicative order of $2$ modulo $N-1$ is the smallest integer $r$ such 
that $2^r=1\,\mbox{mod}(N-1)$,  which is the quantum period again as $S^r=1$.
We are guaranteed that such a number exists as Euler's generalization of 
Fermat's little theorem implies that $\phi(N-1)$ is such that $2^{\phi(N-1)} 
\,\equiv 1\,\, \mbox{mod}(N-1)$, thus $r$ is
either $\phi(N-1)$ or is a divisor of it (here $\phi(n)$ is the Euler totient 
function, being the number of positive integers less than $n$ and coprime to it).
  Finding the multiplicative order is the route of the
  quantum factoring algorithm of Shor. Thus it is interesting that this 
well-known quantum algorithm  makes critical use of an operator that could be 
thought of as a quantization of the fully  chaotic left-shift, or at least 
nearly, as explained below.  
 
\begin{figure}
\includegraphics[width=4in]{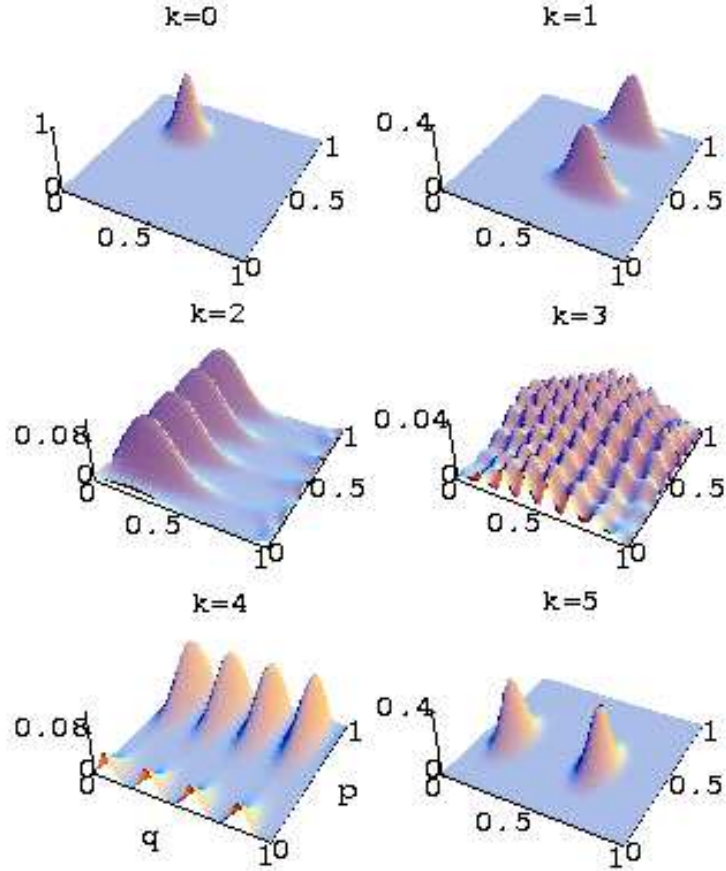}
\caption{The correlation $|\br qp|S^k|q_0,p_0\kt|^2$ as function of $(q,p)$ for the
case of $N=64$, where $|qp\kt$ is a toral coherent state localized at $(q,p)$ \cite{Saraceno}.
 On further applying $S$ to the last figure
 produces the first as in this case $S^{6}$ is the identity. }
\label{sit}
\end{figure}

 That the classical limit of the unitary operator $S$ is not the
 baker's map is made clear by studying its action on coherent states.
 The structure of $S$ in the position basis is that of a permutation, and its
 action on the momentum basis is found easily:
 \beq
 \br m'|S|m\kt = \df{1}{N} \df{-\sin\left[\pi(m'+1/2)/N\right]+
 (-1)^{m+1} \cos\left[\pi (m'+1/2)/N\right]}{\sin\left[\pi (m-2m'-1/2)/N\right]}
 \eeq
 Thus the momentum representation is also real. More importantly for a given 
initial momentum $m$, there are two momentum values around which the final state 
is spread, namely $[m/2]$ or $[m/2] \pm N/2$. Thus the action of $S$ on coherent 
states would be  roughly a combination of its actions on position and momentum 
states, and therefore splits an initial state while performing appropriate 
scaling. Thus $S$ creates ``squeezed cat states'' out of coherent ones, taking a 
state localized at  $(q,p)$ to two that are localized at $(2q \, 
\mbox{mod}\,1,p/2)$ and   $(2q \, \mbox{mod}\,1,(p+1)/2).$ Repeated action by 
$S$ on an initial coherent state is illustrated in Fig.~(\ref{sit}), and exact
revival occurs for the same reason that a deck of cards under the perfect 
riffle-shuffle reorders. 
 
 Using the action of $S$ we can construct a quantum baker's map. The action of 
 choosing the left or right vertical partition is done by the projectors
 $P_1$ and $P_2=I_N-P_1$, where 
\beq
 P_1=\left( \begin{array}{cc}I_{N/2}&0\\0&0 \end{array} \right).
\eeq
 The action of stretching and compression is implemented by $S$, which however 
produces an extra copy, shifted in momentum by one-half. Thus this is in the other
 horizontal partition that divides 
momentum into two equal halves. Thus we once again use projectors, now in 
momentum space to  excise the extra copy and complete the action.
  The full quantum baker built around $S$ is then written as:
 \beq
 B_S= \sqrt{2} \, G_N ^{-1} \left( P_1 G_N S P_1\, +\,P_2 G_N S P_2\right).
 \eeq  
The factor of $\sqrt{2}$ is essential to restore unitarity after the projecting 
actions. This is not yet another quantum baker's map since closer inspection 
shows that it is indeed very close to the usual baker's map in 
Eq.(\ref{bvsbak}). This is seen on rewriting $B_S$ as:
\beq
\label{newB}
B_S= G_N^{-1} \left(\begin{array}{cc}
G_{N/2}^{(\f{1}{2},\f{1}{4})}&0\\0& i\,G_{N/2}^{(\f{1}{2},\f{3}{4})} 
\end{array}\right)\eeq
 
That the usual quantum baker's map is capable of generalizations, including 
arbitrary phases as boundary conditions and relative phases between the two 
blocks in the mixed representation is well-known \cite{BalVor}, though not all of these 
``decorated'' baker's respect the symmetries of parity and time-reversal.
 The operator $B_S$ however shows
the explicit relationship between a quantum baker's map and the solvable operator $S$,
 whose action on the position basis is practically the doubling map restricted to the integers. 
It may be emphasized that even in $B_S$ we are using anti-periodic boundary 
conditions, the phases of $1/4$ and $3/4$ in the $G_{N/2}$ blocks (as well as 
the factor of $i=\sqrt{-1}$) is a direct consequence of the primitive structure 
in Eq.~(\ref{newB}). That the operator obeys parity symmetry follows from the 
fact that $R_N$ commutes with $G_N$ and on verifying that \beq
R_{N/2} G_{N/2}^{(\f{1}{2},\f{1}{4})} = i\, G_{N/2}^{(\f{1}{2},\f{3}{4})} 
R_{N/2}.\eeq

However it does not obey the time-reversal symmetry obeyed by the usual
quantum baker's map. This follows from the preferential treatment of the 
position basis, in which $S$ is a permutation, whereas in the momentum basis it
is not. In the following we use $S$ as an intermediate operator towards 
simplifying states of the usual baker's map $B$ of Eq.~(\ref{bvsbak}). While 
doing so we will also compare the case of the operator $B_S$ wherein there is a 
more explicit relationship; however a more detailed study of the spectra of 
$B_S$ and related operators is itself postponed. 

The operator $S$ is easily diagonalized. The case
 $N=2^K$ is particularly simple, as one sees from the cyclic shifting that $S^K=I_N$, and therefore
 the possible eigenvalues are $\om^l$ where $\om_{K}=e^{2\pi i/K}$, and $0 \le l \le  K-1$.
  The complete set of eigenfunctions
 can be constructed based on the periodic orbits of the full binary left shift.
 When $K$ is composite, an arbitrary $K$-tuple may not produce
 (on action by $S$) an invariant subspace of full dimensionality $K$. Let the number of primitive
 periodic orbits of period $n$ of the left shift map be denoted as $p(n)$, this is 
 the number of primitive binary $n$-tuples, where a primitive $n$-tuple is one that is not 
 a repetition of a shorter string.  If $K$ has divisors
 $d_1,\,d_2, \ldots, d_M$ (including $1$ and $K$), dimensionality of the invariant subspaces
 are $d_i$, and there are $p(d_i)$ of them. In these subspaces the eigenfunctions
  maybe written as
 \beq 
  |\phi_l\kt = \f{1}{\sqrt{d_i}} \sum_{m=0}^{d_i-1}  \om_{d_i}^{lm}\, S^m |\overline{
  a_{d_{i}-1} a_{d_{i}-2}\ldots a_0} \kt.
\eeq
     
\begin{figure}\includegraphics[width=2.5in,angle=-90]{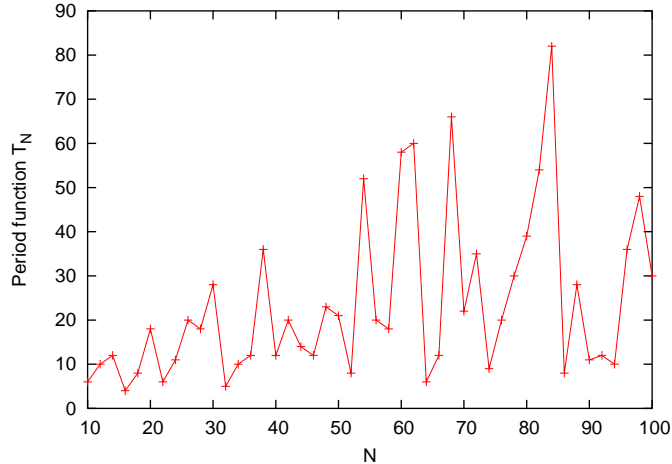}
\caption{The quantum period function $T_N$ which is the multiplicative order
of $2$ modulo $N-1$.}
\label{order}
\end{figure}

The corresponding eigenvalues $\om_{d_i}^{-l}$ are $p(d_i)$-fold degenerate.
 The number of primitive orbits is $p(n)=\sum_{k|n}\mu\left(n/k\right)\, 2^k/n$, 
where $\mu(n)$ is the Mobi\"us function and the sum is over all the divisors of 
$n$. A particularly simple case is when $K$ is prime, as only $d_1=1$ and$d_2=K$ 
are the possible dimensions, and the states in the latter subspace have a 
degeneracy of $(2^K-2)/K=(N-2)/K$. Even when $K$ is not prime the degeneracy 
increases in the same manner for large $N$.
When $N$ is not a power of $2$, the matrix $S$ has nontrivial spectral 
properties. Since$S^{t}|n\kt = |2^t n\, \mbox{mod}(N-1)\kt$,
there exists a time $T_N$ such that $S^T_N=I_N$. This must be the least integer 
such that$2^T \,\equiv 1\,\, \mbox{mod}(N-1)$,  the ``quantum period function'' 
$T_N$ is then simply the multiplicative order defined above, 
$T_N=\mbox{ord}_{N-1}(2)$. This is not a simple function and its solution is 
equivalent to the difficult discrete logarithm problem, and thence to the task 
of factoring numbers.It oscillates wildly with $N$, as seen in Fig.~\ref{order} 
going all the way from $\ln(N)/\ln(2)$ when $N$ is a power of $2$ to $\phi(N-1) 
\sim (N-1) e^{-\gamma}/\ln(\ln(N-1))$, where $\gamma$ is the Euler constant. 

The eigenvalues are then $T_N$-th roots of unity, and one set of eigenfunctions
are given by 

\beq
|\phi_r\kt = \dfrac{1}{\sqrt{T_N}}\sum_{n=0}^{T_N-1}\exp\left(\dfrac{-2 \pi i r 
n }{T_N}\right)\, |2^n \, 
\mbox{mod}\, (N-1) \kt,
\eeq

where $0\le r \le T_N-1$.
For certain $N$ the period $T_N$ is maximal, that is $T_N=\phi(N-1)=N-2$.
 Naturally a necessary condition for this is that $N-1$ be prime. In this case 
 apart from the eigenstates with unit eigenvalues, $|0\kt$ and $|N-1\kt$, the 
others are exhaustively given by the above set. If $T_N \ne N-2$, other 
eigenfunctions can be found based on other subgroups.  
In general
there is degeneracy and the states reside in some appropriate subspace. 

If we use the eigenstates of $S$  as a basis for the eigenstates of the quantum 
baker's map, $B$, or $B_S$ we find remarkable simplifications, as indeed these 
operators are ``close" to each other. The crucial difference is that we can 
solve for the spectrum of $S$ exactly. There are evident similarities of $S$ to 
the well-studied quantum cat maps \cite{cat}, where there is a quantum period 
function that is wildly oscillatory, exactly solvable eigenstates 
\cite{bruno}and so on. Here however the mathematics is far simpler, involving as 
it does a scalar multiplier (namely $2$) rather than an integer $2\times2$ 
matrix. 

\begin{figure}
\includegraphics[width=4in,angle=-90]{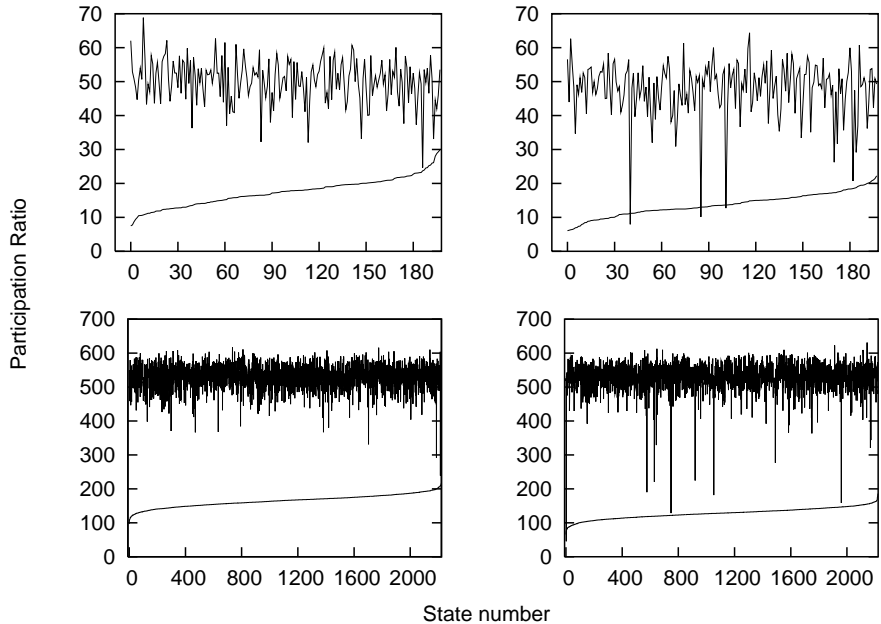}
\caption{The participation ratio in the position and $S$-basis of the quantum 
baker's maps $B$ (left) and $B_S$ (right) when $N=198$ (top) and $N=2222$ (bottom).
These cases are such that $T_N=N-2$. In all the figures the lower curve corresponds to
the $S$-basis, while the upper one to the position basis. The states are arranged
in the increasing order of the participation ratio in the $S$-basis.}
\label{PR}
\end{figure}

Let the eigenvectors of $S$ be $|\phi_r\kt$, we then evaluate the 
participation ratio (PR) $1/(\sum_{r}|\br \phi_r |\psi\kt|^4)$, which gives us 
(roughly) the number of $S$ eigenstates needed to construct the vector 
$|\psi\kt$, here chosen to be one of the eigenstates  of $B$. This is the PR in 
the $S$-basis, while the PR in the position basis is  similarly defined and 
indicates the delocalization in position. For complex random states random 
matrix theory predicts a PR of $N/2$. In Fig.~(\ref{PR}) we compare the 
participation ratio of the eigenstates of $B$ and $B_S$ in both the position and the 
$S$-basis for a particular case, when the the $S$ spectrum is largely 
non-degenerate. The PR in the position basis is halved to take into account the 
parity symmetry of the eigenstates, the $S$-basis already having this symmetry. 
We notice that the $S$-basis ``simplifies'' the states significantly as the PR 
is lesser by a factor of about five or more.

\begin{figure}
\includegraphics[width=4in,angle=-90]{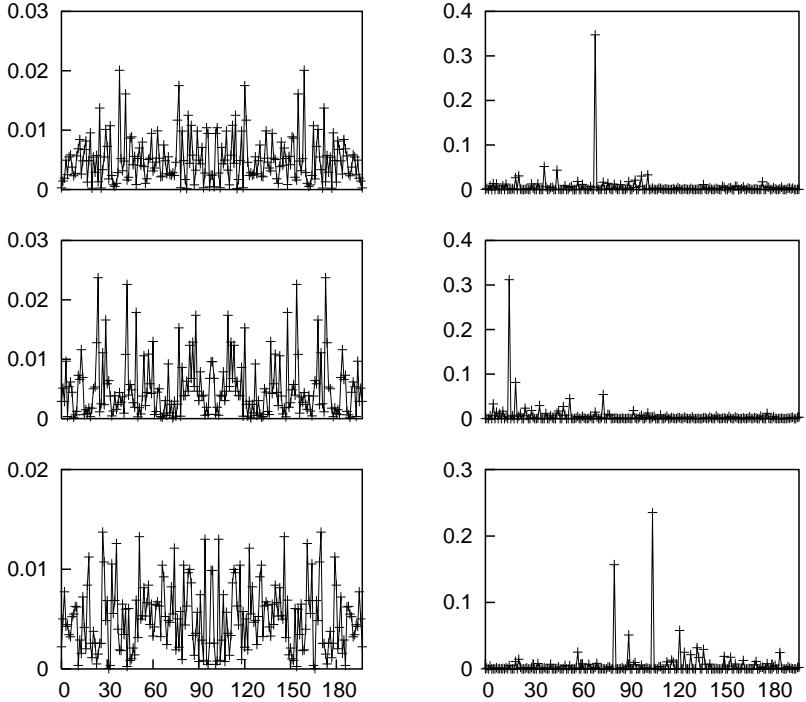}
\caption{The intensity of three eigenstates of the quantum baker's map ($N=198$) 
shown in both the position (left) and in the $S$-basis (right). The states were 
chosen for their contrast in the two basis. }\label{efs}
\end{figure}

We see from the figure that the $S$-basis simplifies states significantly more
in the case of the operator $B_S$, rather than the usual quantum baker's map. 
At the same time, large dips are seen for the eigenstates of $B_S$ that are 
not visible for $B$, indicating perhaps that deviations from RMT (Random Matrix Theory)
are larger in the case of the spectra of $B_S$.    
To illustrate the simplification, we show in Fig.~(\ref{efs}) three eigenstates 
of the usual quantum baker's map $B$, for the case 
$N=198$ that are considerably simplified in the $S$-basis. 

We may improve upon the $S$-basis by making it compliant with time-reversal 
symmetry. For instance, in the first state (say $|\psi\kt$) shown in 
Fig.~(\ref{efs}), the maximum overlap with an $S$-eigenstate $|\phi_r\kt$ is 
$|\br \phi_r|\psi\kt|^2 =0.34$, while the (unnormalized) adapted state 
$|\phi_r'\kt = |\phi_r\kt + G_N^{-1} |\phi_r\kt^*$ has an overlap of $0.37$. 
This adapted state is such that $G_N |\phi_r'\kt = |\phi_r'\kt ^*$ as required 
by time-reversal invariance of the quantum baker's map. An arbitrary phase 
between $|\phi_r\kt $ and $G_N^{-1} |\phi_r\kt^*$ was set as zero after 
numerically ascertaining that this was the optimal value.  Note that the 
conjugation assume that the states are in the postion representation.  

We remark that this simplification falls significantly short of that achieved by 
the Hadamard basis for the case when $N$ is a power of $2$ \cite{meen}. In this 
case (for the operator $B$) the Thue-Morse states and many others simplified
 considerably more in the 
Hadamard basis, or after a Walsh-Hadamard transform; for instance in the case 
when $N=1024$, after the transform the participation ratio of the Thue-Morse 
state was of the order of $2$. While the Thue-Morse sequence (rather its finite 
truncations) is an eigenstate of $S$, the Hadamard transform itself commutes 
with $S$. Due to the degeneracy in the spectrum of $S$, it appears that the 
Hadamard transform represents a basis that is more optimal than that provided by 
the eigenvectors of $S$. The meaning of this commutation of $S$ and $H$ 
perhaps in terms of a classical symmetry is not clear to the author. 

Finally we remark on the statistical properties of the eigenstates, and on 
the ``relative randomness'', in the sense of Kus and Zyczkowski \cite{Kus}, of $S$
and the operators $B$ and $B_S$. The usual quantum baker's map eigenstates
are nearly generic in the sense that they are close to those that are
expected from RMT \cite{Haake}, however there are also known and 
significant deviations, whose origins may be number-theoretic (such as the multifractal
scaling of eigenstates for the case when $N$ are powers of $2$ \cite{meen}).
We find, from results not presented here, that while the eigenstates in the position
basis are much closer to the expected Porter-Thomas distribution, the eigenstates
in the $S$-basis are considerably deviated, as is to be expected. 

\begin{figure}
\includegraphics[width=2.5in,angle=-90]{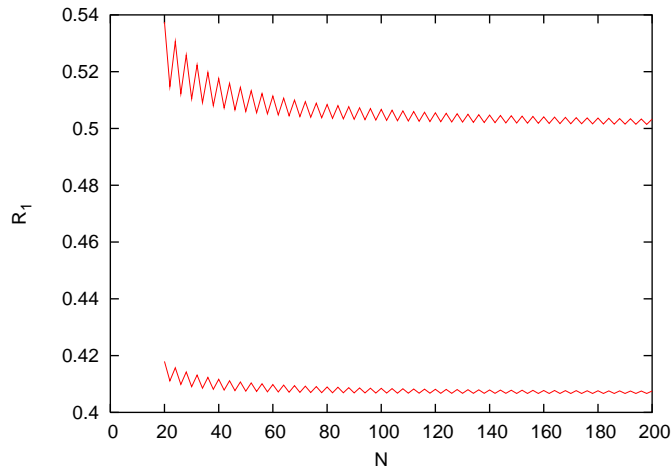}
\caption{The relative randomness measure $R_1$ as function of $N$, between the 
operators $S$ and $B_S$ (upper curve), and between $S$ and $B$ (lower curve).}
\label{relrand}
\end{figure}

To quantitatively compare $S$ and the baker's map operators
$B$ and $B_S$, we study their relative randomness, or degree of 
noncommutativity, by means of the inner-product between
the operator $S$ and its image under $B$ (or $B_S$). Thus define 
\beq
R_1=|\br S|BSB^{\dagger} \kt|/N,
\eeq
where $\br X |Y \kt = \mbox{Tr}(X \, Y^{\dagger})$. It is argued in \cite{Kus} that 
this (and related quantities) are small, near zero, if the operators $S$ and
$B$ are relatively random, whereas if they commute or anticommute $R_1=1$.
We show in Fig.~(\ref{relrand}) this measure for both the operators $B$ and $B_S$
as a function of $N$. It is clear that the quantum baker's map $B$ is significantly
correlated to the operator $S$, as the inner-product $R_1$ is around $0.4$,
 and that the operator $B_S$ is more so correlated,
 as the inner-product is around $0.5$. This is of course reflected in 
the fact that the eigenstates of $B_S$ are more compressed in the $S$-basis. It is 
worthwhile remarking that powers of $2$ do not appear to be special for the measure $R_1$.
Also the inner-products between $S$ and the baker's map operators themselves behave similarly,
as $|\br S|B\kt |/N \approx 0.63$ while  $|\br S|B_S\kt |/N \approx 0.70$.

In conclusion the exactly solvable operator $S$ is a good ``zeroth order'' 
system for the quantum baker's map. This operator is somewhat similar to the 
semiquantum operators that are obtained on quantizing classical baker's after 
times larger than one \cite{Almeida}. However these operators usually have 
complicated spectra themselves. We can use $S$ to build a quantum baker's map,
which is very close to the usual baker's map, which in turn explains the
close relationship between the solvable spectrum of $S$ and that of the 
quantum baker's map. Using a relative randomness measure it has
been shown that indeed the operator $S$ is significantly correlated with the
quantum baker's map. While  pointing to the evident connection of $S$ to the
 task of factoring numbers, it is tempting to speculate  that the relationship
  between classically hard computations and their (probably faster) 
  quantum algorithms has a deeper connection to the transition from classical
   to quantum chaos.    
 
\acknowledgments{I thank the referee for useful suggestions and pointing to 
reference \cite{Kus}.}

\end{document}